# Crisis Alpha: A High-Performance Trading Algorithm Tested in Market Downturns

**Maysam Khodayari Gharanchaei [1*], Reza Babazadeh[2,†]**

[1]*Tepper School of Business, Carnegie Mellon University, Pittsburgh, Pennsylvania, US*

[2] *Faculty of Engineering, Urmia University, Urmia, West Azerbaijan Province, Iran*

**Abstract**

Forming quantitative portfolios using statistical risk models presents a significant challenge for hedge funds and portfolio managers. This research investigates three distinct statistical risk models to construct quantitative portfolios of 1,000 floating stocks in the US market. Utilizing five different investment strategies, these models are tested across four periods, encompassing the last three major financial crises: The Dot Com Bubble, Global Financial Crisis, and Covid-19 market downturn. Backtests leverage the CRSP dataset from January 1990 through December 2023. The results demonstrate that the proposed models consistently outperformed market excess returns across all periods. These findings suggest that the developed risk models can serve as valuable tools for asset managers, aiding in strategic decision-making and risk management in various economic conditions.

## 1. Introduction

The portfolio quantitative risk management concept has been a mainstay of asset management practice after consecutive financial crises throughout the world's capital markets. On the other hand, portfolio managers have had concerns over the efficiency of investment diversification strategies equipped with appropriate risk models. While applying risk models and diversification strategies have been the pivot points of portfolio management in financial institutions, portfolio managers face severe challenges in applying them during financial crises. This is because regime changes causing the crises. The regime changes occur when the expectations of market stakeholders are changed drastically about the future scenarios in markets. This is the challenge which risk model would face to overcome since these models need to be updated on a massive data points of time series which is computationally expensive. According to (Shiller, 2015) many economic bubbles occur when investors assume that a new era has begun in the economy. These "new eras" are based on the belief that a particular trend or factor has permanently changed the market, even though the market structure itself remains the same that emphasis to use an objective procedure to control the risk of investment and implement a proper diversification on investment holdings. Investment funds consider expected market scenarios (subjective element) and structural correlations of market factors (objective element) Khodayari Gharanchaei (2024a, 2024b, 2024c); Khodayari et al. (2019). The former is a trading strategy, and the latter is a risk model or covariance matrix of assets. Also, Prasad Panda et al. (2024) used neural network model dealing with many uncertainties for factor timing in asset management.

[*] First Corresponding author: M. Khodayari G. (mkhodaya@andrew.cmu.edu)
[†] Second Corresponding author: R. Babazadeh (r.babazadeh@urmia.ac.ir)

The dot com bubble was a period of excessive speculation in internet-related companies. The Nasdaq Composite Index soared by almost 400% over the five years from 1995 to 2000. According to (Burks et al., 2021), "in the two-year period from early 1998 through February 2000, the Internet sector earned over 1000 percent returns on its public equity". However, once this bubble burst, the Nasdaq lost 78% of its value. The Global Financial Crisis (GFC) of 2008 was the first major economic crisis to hit the world since the Great Depression. It originated from the US subprime mortgage market and had a significant impact on the global economy, leading to a decline in demand and financial liquidity. Moreover, the highly interconnected nature of the markets played a crucial role in the spread of the crisis (Silva et al., 2023). According to (Zhang & Broadstock, 2020) global economic growth suffered a loss of over $2 trillion during the crisis and between 2007 and 2010, around 6 million homes in America were foreclosed. The crisis caused major declines in stock markets globally. For instance, the S&P 500, a benchmark for U.S. equities, fell by approximately 57% from its high in October 2007 to its low in March 2009 (Wang et al., 2017). Hence, the importance of large-scale correlation-based quantitative risk models was inevitable after the GFC. For this reason many papers have shown the importance of the determinants of the GFC and their correlations such as (Luchtenberg & Vu, 2015). Unlike any previous financial crises, the COVID-19 pandemic caused a one-of-a-kind financial crisis in 2020, leading to a global economic shutdown. The stock markets witnessed significant falls due to the uncertainty surrounding the pandemic's impact on the global economy. According to (*World Bank*, 2020) global GDP was projected to shrink by 3% in 2020, the steepest slowdown since the Great

Depression. The pandemic caused a significant rise in unemployment rates worldwide, with the U.S. unemployment rate reaching 13.0% in the second quarter of 2020 (Smith et al., 2021).

Several types of quantitative risk models are commonly used in stock portfolio management. Value at risk (VaR) and conditional value at risk (CVaR), also known as expected shortfall, are the basic statistical risk models that rely on the cumulative distribution function of expected losses. According to (Happersberger et al., 2020) the VaR does not have subadditivity property for all types of loss distributions and while the CVaR is a subadditive risk measure, it is used as an informative metric of the expected loss of a portfolio beyond a specific confidence level. The Monte Carlo Simulation (Stevens, 2022) method involves repeated random sampling to simulate a range of potential portfolio outcomes under varying scenarios. It can help evaluate the potential impact of risk under different conditions (Mukherjee et al., 2022). After the GFC, factor risk models have been used widely by researchers and industry pioneers because they can estimate the correlation among the stocks of a portfolio (Ang et al., 2020). A factor risk model is a tool used by investors to estimate the risk and relationship between securities. This model allows investors to construct the covariance matrix of the assets in the portfolio (So et al., 2022). While all of these models use regression on single or multiple independent factors (De Nard et al., 2019), they are divided in two categories: $i$) Time-series and $ii$) Cross-sectional risk models. The Time-series model uses independent factor returns like market excess return, size, and value of securities to predict the expected returns of assets but the Cross-sectional model uses returns of the portfolio's securities over a single time period to predict the expected returns of the portfolio. Both methods estimate regression loadings as inputs to construct the covariance matrix of assets' returns (Fama & French, 2019). Also, researchers have used principal component analysis (PCA) based risk models since emerging advanced computational machines, enabling them to run heavy algorithms. According to (Mavungu, 2023) these models use PCA to compute and extract the main factors for the risk of a portfolio, to determine the most dominating stock for each risk factor and each portfolio, and finally, to compute the portfolio's total risk.

In this research paper, we have compared the outcomes of three different types of risk models. We used the time-series factor risk model, constant correlation covariance matrix, and sample covariance matrix with shrinkage to create the covariance matrix of the assets in the portfolio. This covariance matrix is an argument of the objective function in quadratic optimization problems for investment diversification strategies. The risk models and diversification strategies are used to construct portfolios based on the out-of-sample data. Finally, we compared the results of the out-of-sample tests over the last three financial crises in the 21st century. Using the time-series factor risk model, we can perform a regression analysis to determine the length of time between two consecutive financial crises. This allows the risk model to capture important information during these periods. However, the correlation among traded securities is often disrupted during a crisis due to high volatility and market perturbations. This can lead to crashes and price drops in many stocks, including those that were previously profitable and underpriced, as they tend to show similar behavior during these times. This pattern can be modeled through constructing a constant correlation covariance matrix over the portfolio holdings. For this purpose and also to decrease the sensitivity of the covariance matrix to its inputs, we used the shrinkage method developed by Clarke et al. (2006) on the sample covariance matrix.

The main contributions of this paper that differentiates it from the available works in the literature include:

- Developing statistical risk models using statistical attributes among hundreds of stocks and idiosyncratic characteristics of individual assets;
- Leveraging quadratic numerical optimization techniques;
- Utilizing a new shrinkage method to control the covariance risk model's sensitivity;
- Transforming the risk parity objective function in the optimization phase to create a convex quadratic optimization problem;
- Applying the proposed approach to a real dataset by using the CRSP dataset to perform a simulated back-test.

The structure of this paper is as follows: Section 2 reviews recent significant research papers to identify trends in the field of quantitative risk management. Section 3 provides a comprehensive overview of quantitative risk models, presenting three specific models along with their mathematical formulations. This section also discusses quadratic optimization frameworks for various diversification strategies. Section 4 details the results of a back-test simulation using CRSP datasets across different time periods corresponding to three financial crises. The analysis of back-test results aims to determine which risk model or investment strategy demonstrated superior performance across these periods. Managerial implications are provided at the conclusion of Section 4, serving as a professional guideline for researchers in both academia and industry.

Finally, Section 5 summarizes the framework and findings of this research, offering suggestions for future research avenues for interested scholars.

## 2. Literature Review

In recent years, many decent research works are done in the field of risk management and portfolio optimization. Notable papers are published in which alternative datasets are used through AI equipped methods such as deep learning natural language processing (Ding et al., 2015). Also many hybrid works have leveraged both classic big datasets and alternative datasets to generate high yields while controlling portfolio risk (In et al., 2019). When it comes to investing in stocks, selecting the right stocks is crucial. This usually involves evaluating both the potential return and the associated risk. The standard approach for assessing risk and return is to use the Sharpe ratio. However, the Sharpe ratio uses the standard deviation to measure portfolio risk, which can be problematic because a portfolio with an upward trend can still have high risk, even though this contradicts the common logic of most investors. Trend Ratio (TR), a novel assessment strategy, solves this problem (Chou et al., 2018). TR uses linear regression, which is improved with the initial funds when seeking to find the trend of a portfolio.

In terms of detective risk models, de Castro Vieira et al. (2019) investigated the behavior of default prediction models based on credit scoring methods and computational techniques with machine learning algorithms and concluded that the best prediction results were obtained with traditional ensemble techniques — in this case Bagging, Random Forest, and Boosting. Factor investing has been a highly attractive subject to do computationally expensive researches in the last decade especially after the GFC. In their study, Zhao et al. (2019) investigated the predictability of the Fama-French five factors and their optimal allocation in factor investing from 2000 to 2017. The research employs a combination of linear and nonlinear models for forecasting, enhanced by dynamic model averaging, and compares these forecasts with various other prediction techniques. It also utilizes the generalized autoregressive score model and the skewed t copula method for estimating asset correlation, offering an innovative approach to forecasting covariance and improving portfolio diversification. The findings suggest significant advancements in portfolio optimization by integrating Bayesian forecast combinations and copula methods, particularly in managing asymmetric dependencies among factors. The application of advanced numerical methods has been expanding not only in developing risk models but also in forming portfolio diversification strategies.

Molyboga (2020) introduced a Modified Hierarchical Risk Parity (MHRP) approach, enhancing traditional HRP by integrating three key elements: an exponentially weighted covariance matrix with Ledoit-Wolf shrinkage, an equal volatility allocation method, and volatility targeting for improved diversification. His large-scale Monte-Carlo simulation, focusing on Commodity Trading Advisors, reveals a notable 50% improvement in the out-of-sample Sharpe ratio and a reduction in downside risk, demonstrating the efficacy of MHRP in addressing the limitations of conventional risk parity models. According to (Pedersen et al., 2021) standard mean–variance optimization works poorly in practice that optimization is often abandoned. They generated significant alpha beyond the market through their novel portfolio optimization method that has a correlation shrinkage parameter, which is chosen to maximize risk-adjusted returns in past data. Al Janabi (2021) examined the theoretical foundations for multivariate portfolio optimization algorithms under illiquid market conditions. They showed that portfolio and risk managers are able to set various closeout horizons and dependence measures to compute the required Liquidity-adjusted Value at Risk (LVaR) and determine the resulting investable portfolios. Additionally, portfolio managers can compare the return-to-risk ratio and asset allocation of the acquired investable portfolios with different liquidation horizons as compared to the conventional Markowitz mean-variance approach. Chen et al. (2021) through data envelopment analysis (DEA) introduced a novel portfolio optimization framework by incorporating environmental, social, and governance (ESG) performance. They could define ESG scores with financial indicators to select assets based on a cross-efficiency analysis.

After the pandemic, due to supply chain disruption and its multi-aspects effects, researchers in quantitative finance field have focused on developing financial models and risk management strategies for portfolios that include both stocks, crypto and derivatives. Wang et al. (2022) enhanced the return of a portfolio of five energy futures from January 2011 to July 2020 by taking short positions in the same level of risk and applying same risk models. Díaz et al. (2022) explored the diversification benefits of socially responsible investments (SRIs) during the COVID-19 pandemic. The research focused on the impact of integrating clean energy equities into portfolios containing conventional equities and traditional safe-haven assets. Utilizing AR-GARCH models and copula specifications, the study demonstrated SRIs' effectiveness in enhancing portfolio performance and risk diversification, particularly against a backdrop of varied assets including equities, bonds, gold, oil, and Bitcoin. With emerging advanced methods of high-scale optimization problems, the need for performing optimal shrinkage is more significant than ever.

Bodnar et al. (2022) presented a novel linear shrinkage estimator for mean-variance portfolio optimization in high-dimensional scenarios, utilizing advancements in random matrix theory. This estimator, notable for incorporating estimation risk and being distribution-free, optimizes the asymptotic out-of-sample expected utility across various risk aversion levels. Demonstrating superior performance over existing methods, especially in cases where portfolio dimension exceeds sample size, the study also confirms the estimator's robustness against non-normality and its efficacy through both numerical and empirical analyses under minimal assumptions about asset return distributions. Zha et al. (2023) examined the dynamic relationships and risk spillovers among Bitcoin, crude oil, and various traditional markets (U.S. and Chinese stocks, gold, bonds, currency, and real estate) during the COVID-19 pandemic. Utilizing quantile-on-quantile, time-varying copula, and conditional value-at-risk models, the study revealed significant, pandemic-induced alterations in market interconnections, with Bitcoin notably impacting the crude oil market and others. This research highlights the necessity for investors and policymakers to acknowledge these risk spillovers and adapt their strategies accordingly in the face of economic challenges posed by such global crises. Owen (2023) explored portfolio optimization by enhancing ex-ante conditional covariance estimates using hierarchical clustering, a machine learning algorithm. The research, utilizing a 52-year dataset of stock returns, demonstrated that this method outperformed traditional Markowitz portfolios in terms of risk-adjusted returns when applied in a 3-month buy-and-hold, long-only strategy. Notably, it also resulted in significantly lower portfolio weight changes at each rebalancing period, effectively reducing trading costs. The findings remained robust across various settings and subsamples. Not only have exotic portfolios been an interesting research domain after the GFC, but AI-equipped algorithms have also been an area of interest in computational research projects. Ajami (2024); Ajami et al. (2023); Nigjeh et al. (2023).

Ma et al. (2024) introduced a novel risk and return multi-task learning model, the HGA-MT, which employs a Heterogeneous Graph Attention (Multi-Task) network for enhanced stock ranking prediction in portfolio optimization. This model uniquely integrates graph convolutional networks and an attention mechanism to effectively aggregate and weigh multiple spillover effects from related stocks, addressing both stock return and volatility risks. The HGA-MT's approach of jointly learning stock return and volatility risks marks a significant advancement in accurately identifying top-ranked stocks for portfolios. The model demonstrated superior performance over existing methods in stock ranking and back-testing trading evaluations, suggesting its efficacy for achieving higher and more stable profits. Alzaman (2024) explores the integration of deep learning (DL) with genetic algorithms for portfolio selection in financial markets. The research demonstrates a 40% improvement in prediction accuracy over traditional grid searches by employing genetic-based hyper parameter optimization. This approach, combined with multiple Deep RankNet models, notably outperformed general market returns by 20%, and more than doubled during the volatile COVID-19 period. The study represents a significant advancement in leveraging deep learning and genetic algorithms for enhanced financial asset prediction and portfolio management. Finally, the individual behavioral preferences of investors are a novel area to work on through quantitative frameworks. Guo et al. (2024) innovatively integrates fuzzy preference relations (FPRs) with Markowitz portfolio theory, addressing the portfolio selection challenge under uncertain, fuzzy, and incomplete investor preferences. The research improves the relationship between judgment elements and priority vectors in FPRs, proposing novel portfolio models for both unknown and partially known preference scenarios. Through empirical analysis, the study reveals the positive impact of increasing the distance threshold in FPR-based models and demonstrates their robustness to consistency variations in FPRs. This work marks a significant advancement in blending behavioral aspects with quantitative portfolio construction.

## 3. Proposed Methodology

This study employed the CRSP dataset, which includes risky US stocks, using adjusted returns to neutralize dividend effects. We evaluated four investment diversification strategies using 1000 largest stocks of US markets across three risk models and four periods, including dot-com bubble, 2008 global financial crisis, COVID-19 crisis, and post-COVID period through December 2023. Monthly timeseries data, consisting of the past 60 monthly excess returns and market excess returns, were utilized to estimate the covariance matrix and optimal holdings for each scenario. The market returns and risk-free returns were downloaded from Kenneth French's data library.

Three independent risk models were employed: single-factor covariance matrix, constant correlation covariance matrix, and sample covariance matrix with shrinkage.

In the first risk model, in equation (1), $\sigma_f^2$ serves as the variance of market returns, the sole factor of the risk model. The vector $\boldsymbol{b}$ represents the adjusted factor loadings for $n$ assets, and $\boldsymbol{D}$ is the adjusted diagonal matrix of idiosyncratic asset volatilities. Adjustment of **b** and **D** utilizes formulas (2) and (3) for shrinkage.

$$V = \sigma_f^2 \cdot bb^T + D \quad (1)$$

$$\beta_i := \hat{\beta}_i + \frac{1}{3} \cdot (1 - \hat{\beta}_i) = \frac{2}{3} \cdot \hat{\beta}_i + \frac{1}{3} \quad (2)$$

$$\log \omega_i := \log \hat{\omega}_i + \frac{1}{3} \cdot \left( \frac{1}{N} \sum_{j=1}^{N} \log \hat{\omega}_j - \log \hat{\omega}_i \right) = \frac{2}{3} \cdot \log \hat{\omega}_i + \frac{1}{3N} \cdot \sum_{j=1}^{N} \log \hat{\omega}_j \quad (3)$$

In the second risk model, covariance matrix is derived through formula below

$$V = \rho \cdot \sigma\sigma^T + (1 - \rho) Diag(\sigma)^2 \quad (4)$$

In which, $\rho$ is the (constant) correlation between any two different stock returns and $\sigma$ is the vector of stock volatilities. To estimate $\rho$ and $\sigma$ we estimated $\rho$ as the average of all sample correlations, $\rho_{ij}$. Also, we estimated each $\sigma_i^2$ via sample variances and then shrink the log volatilities $\frac{1}{3}$ towards their average log volatilities.

For the last risk model, we estimated the covariance matrix by shrinking the sample covariance matrix towards a structured target matrix. The method employed in this model is the same method used by (Clarke et al., 2006) which was used for the US equity market from January 1968 through December 2005. Accordingly, we tested their method across time periods that included the Global Financial Crisis (GFC) and the COVID-19 pandemic.

After constructing the risk models, we formulated three distinct optimization problems to generate minimum variance, maximum diversification, and risk parity portfolios. These optimization problems were solved using the CVXPY library, a Python-based tool that supports various solvers, including ECOS, SCS, and OSQP. Notably, we employed GUROBI as the solver.

Minimum variance portfolio focuses on minimizing the overall risk, quantified as the variance of expected returns. The core objective is to identify a combination of holdings that yields the lowest possible portfolio volatility. These holdings are derived from the optimal solution to the following convex optimization problem.

$$\begin{aligned} \min_{X} \quad & X^T V X \quad (5) \\ s \cdot t \cdot \quad & \mathbf{1}^T X = 1 \\ & X \leq u \\ & x_i \geq 0 \quad i = 1, 2, \cdots, n \end{aligned}$$

Here, **u** is an upper band set to avoid unintentional portfolio concentration on a few stocks. We set it equal to 5% for all assets.

In constructing portfolios, to avoid focusing on risk only, there are several approaches to diversify the assets in a way to gain better portfolio returns while controlling the risk distribution. The maximum diversification portfolio method is a portfolio optimization strategy designed to achieve the highest level of diversification among a set of assets to gain exposure to higher risks versus a greater risk degree of freedom. By assuming

$$KX = Z \quad (6)$$

the maximum diversification optimization problem for a long-only and fully-invested portfolio can be expressed as

$$\min_{Z} \quad Z^T V Z \quad (7)$$
$$s \cdot t \cdot \quad \sigma^T Z = 1$$
$$\mathbf{1}^T Z = K$$
$$Z \leq K\mathbf{u}$$
$$K \geq 0$$
$$z_i \geq 0 \quad i = 1, 2, \cdots, n$$

Here, K is an arbitrary coefficient that satisfies equation (6).

The Risk Parity strategy aims to equalize the risk contribution of each asset within the portfolio, assigning lower weights to higher-volatility assets and higher weights to lower-volatility assets. This approach seeks to balance risk contributions across all holdings, ensuring no single asset dominates the portfolio's risk profile. Unlike earlier optimization approaches, the risk parity model is numerically sensitive due to the logarithmic term in its objective function. To address this, we implemented an upper band set at five, a sufficiently large positive integer to prevent the optimization problem from becoming non-positive definite (NPD). Successfully completing all out-of-sample tests, the risk parity portfolio under a long-only and fully-invested condition can be derived by solving the specified convex optimization problem below.

$$\min_{Y} \quad \frac{1}{2} Y^T V Y - \sum_{i=0}^{n} \log(y_i) \quad (8)$$
$$s \cdot t \cdot \quad Y \geq 0 \quad and \quad Y \leq d$$

Y is defined as

$$X = \frac{1}{\mathbf{1}^T Y} Y \quad (9)$$

We noted that when employing a sample covariance shrinkage model as the basis for the covariance matrix, the positive semi-definite (PSD) condition is not numerically satisfied in the risk parity strategy by the solver. This indicates a computational challenge in maintaining the PSD requirement essential for the optimization's stability and validity. So it is recommended to employ other diversification strategies along with this risk model.

## 4. Computational Results

We conducted out-of-sample tests on 1000 assets across four crisis periods to assess the efficacy of various risk models combined with multiple investment strategies. Given the substantial number of assets, we were able to develop a high-frequency algorithm to construct various risk models and ultimately, investment portfolios. To enhance efficiency and reduce computational costs associated with intensive calculations, instead of various python regression packages, we utilized matrix operations to estimate model elements such as such as $\hat{\beta}_i$, $\hat{\omega}_i^2$, $\hat{\rho}_{ij}$, $\hat{\sigma}_i^2$ and shrinkage calculations. We used the average positions and effective number of stocks (N) as defined by Strongin et al. (2000). According to that research, the effective number of stocks (N) in the portfolio, can be interpreted as the number of stocks that could be equal-weighted to get the same level of stock-specific risk as occurs in the original portfolio.

The results of out-of-sample tests across various financial crises and post-COVID-19 market conditions are displayed in the following tables. Each table corresponds to the application of a risk model over a designated period, with each column representing a specific investment strategy. These strategies employ convex objective functions that utilize the risk model as their argument.

Tables 1(a), 1(b), and 1(c) present the results of back-test simulation from January 1990 through March 2000 ending in Dot Com bubble crisis.

Table 1(a) presents the Market Factor Risk Model results. This model evaluates the performance of five investment strategies: Market (Value weighted), Equal Weighted, Minimum Variance, Maximum diversification, and Risk Parity. The critical metrics analyzed include the average excess return, standard deviation, Sharpe ratio, market beta, average positions, and effective N. As shown below, the maximum diversification strategy exhibited the highest average excess return at 35.32%, whereas the minimum variance strategy showed the lowest at 2.04%. The minimum variance strategy had the lowest standard deviation at 8.85%, indicating lower risk, while maximum diversification had the highest at 19.96%. These results are in line with the mean-variance theory. The maximum diversification strategy also achieved the highest Sharpe ratio of 1.77, suggesting the best risk-adjusted return among the strategies. The value-weighted strategy had a beta of 1.00, as expected (as the indicator of market excess return).

In contrast, the minimum variance strategy had a significantly lower beta of 0.06, reflecting reduced market sensitivity. The equal-weighted and value-weighted strategies maintained high average positions (1000 stocks), while the minimum variance and maximum diversification strategies had significantly fewer positions. These strategies rely on stocks with more influence on the portfolio in terms of lower risk and higher return. The accumulative returns are depicted in Figure 1(a).

**Table 1(a).** Market Factor Risk Model – Dot Com Bubble (01/1990-03/2000)

|  | Market (Value-Weighted) | Equal Weighted | Minimum Variance | Maximum Diversification | Risk Parity |
|---|---|---|---|---|---|
| Average Excess Return | 17.09% | 14.13% | 2.04% | 35.32% | 11.03% |
| Standard Deviation | 13.41% | 14.07% | 8.85% | 19.96% | 11.97% |
| Sharpe ratio | 1.27 | 1.00 | 0.23 | 1.77 | 0.92 |
| Market Beta | 1.00 | 0.99 | 0.06 | 0.48 | 0.82 |
| Average Positions | 1000.0 | 1000.0 | 68.7 | 65.1 | 1000.0 |
| Effective N | 190.2 | 1000.0 | 38.8 | 39.1 | 866.4 |

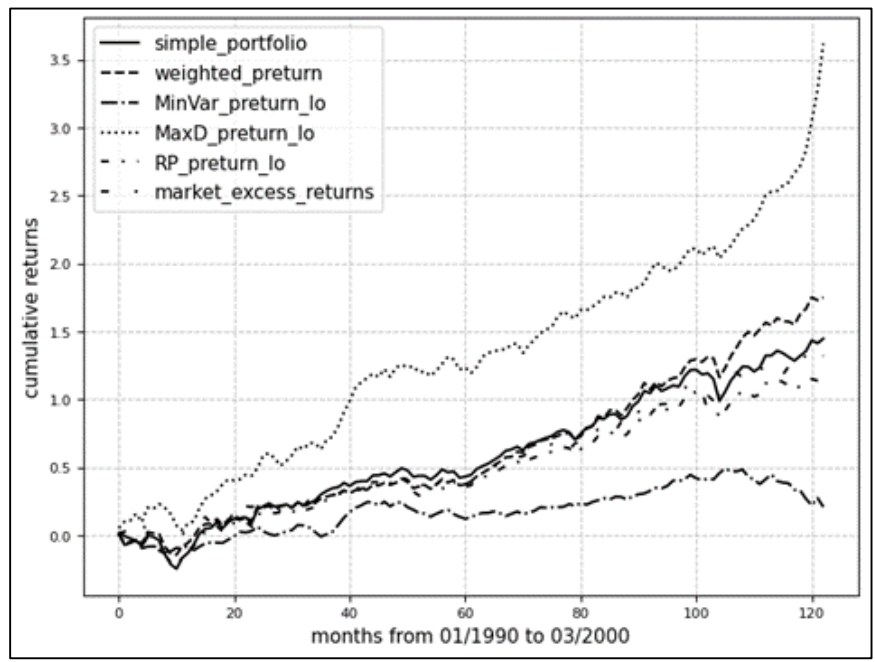

**Figure 1(a).** Market Factor Risk Model – Portfolios Comparison – Dot Com Bubble

Table 1(b) details the results of the Constant Correlation Covariance Matrix model. This model also assesses the same five investment strategies but uses a different approach to covariance estimation. The maximum diversification strategy again showed the highest return at 80.83%, while the minimum variance strategy had a negligible return at -0.02%. As expected, the minimum variance strategy had the lowest standard deviation at 7.53%, and maximum diversification had the highest at 52.27%. The Sharpe ratio was highest for the maximum diversification strategy at 1.55, while the Minimum Variance strategy had a ratio of 0.00, indicating no risk-adjusted return. As a result, the maximum diversification strategy is the most sensitive to the market return by holding a beta of 2.11. Like the previous model, the minimum variance and maximum diversification strategies had lower average positions than the value-weighted and equally weighted portfolios. The accumulative returns are shown in Figure 1(b).

**Table 1(b).** Constant Correlation Covariance Model – Dot Com Bubble (01/1990-03/2000)

|  | Market (Value-Weighted) | Equal Weighted | Minimum Variance | Maximum Diversification | Risk Parity |
|---|---|---|---|---|---|
| Average Excess Return | 17.09% | 14.13% | -0.02% | 80.83% | 13.07% |
| Standard Deviation | 13.41% | 14.07% | 7.53% | 52.27% | 13.66% |
| Sharpe ratio | 1.27 | 1.00 | 0.00 | 1.55 | 0.96 |
| Market Beta | 1.00 | 0.99 | 0.33 | 2.11 | 0.96 |
| Average Positions | 1000.0 | 1000.0 | 21.6 | 20.1 | 1000.0 |
| Effective N | 190.2 | 1000.0 | 20.6 | 20.0 | 993.0 |

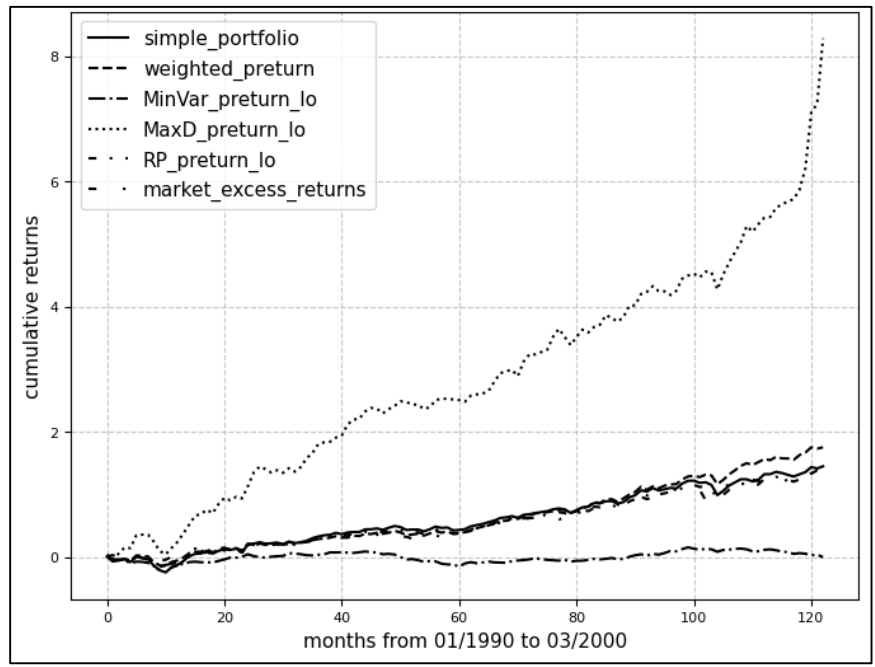

**Figure 1(b).** Constant Correlation Covariance Model – Portfolios Comparison – Dot Com Bubble

Table 1(c) illustrates the results using the Sample Covariance Matrix with the Shrinkage model, providing a more robust estimation by shrinking the sample covariance matrix towards a structured target. Repeatedly, the maximum diversification strategy led to an average excess return of 35.90%. On the other hand, the minimum variance portfolio, as the most conservative strategy, had the lowest standard deviation at 4.96% with the lowest return. As a result, the maximum diversification strategy had the highest Sharpe ratio of 1.89, indicating superior risk-adjusted performance. While advanced numerical optimization packages were utilized to run back-tests, the shrunk covariance model skewed the convexity of the objective function of the risk parity portfolio. In function number (8), a constant value has to be added to either logarithmic or quadratic terms. Because of complex numerical adjustments, this modification could be performed in a separate study to find the best methodology for overcoming this issue. For this paper, the risk parity strategy could not be combined with the shrunk sample covariance risk model during all four timeframes. The accumulative returns are depicted in Figure 1(c).

**Table 1(c).** Sample Covariance Matrix with Shrinkage – Dot Com Bubble (01/1990-03/2000)

|  | Market (Value-Weighted) | Equal Weighted | Minimum Variance | Maximum Diversification | Risk Parity |
|---|---|---|---|---|---|
| Average Excess Return | 17.09% | 14.13% | 2.76% | 35.90% | NaN |
| Standard Deviation | 13.41% | 14.07% | 4.96% | 19.02% | NaN |
| Sharpe ratio | 1.27 | 1.00 | 0.56 | 1.89 | NaN |
| Market Beta | 1.00 | 0.99 | 0.28 | 1.09 | NaN |
| Average Positions | 1000.0 | 1000.0 | 171.3 | 45.3 | NaN |
| Effective N | 190.2 | 1000.0 | 99.0 | 29.1 | NaN |

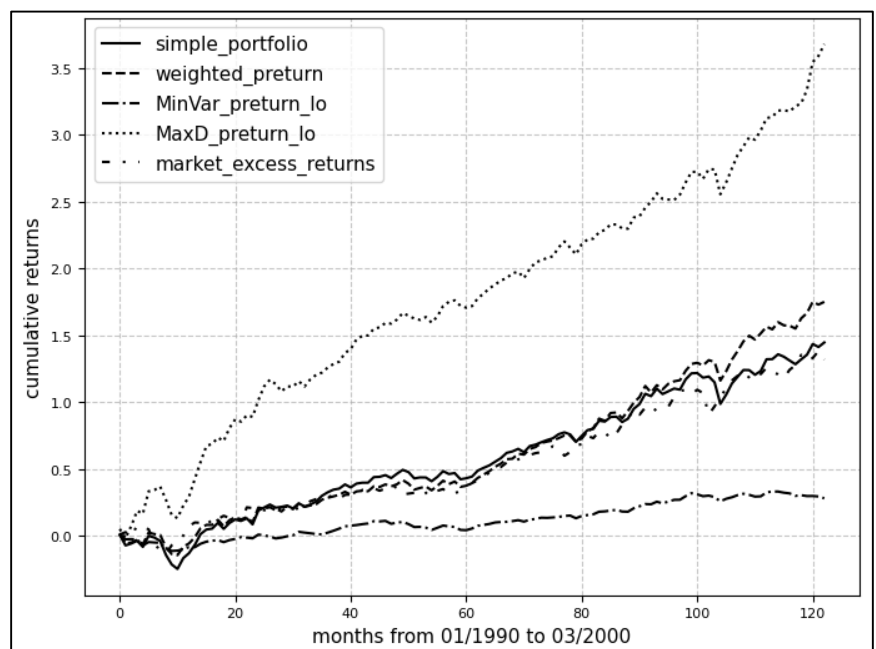

**Figure 1(c).** Shrunk Sample Covariance Matrix – Portfolios Comparison – Dot Com Bubble

During the Dot-Com Bubble period (January 1990 - March 2000), the maximum diversification strategy exhibited the highest average excess returns across all models and the highest Sharpe ratios. These results highlight the strategy's robustness and ability to capitalize on market conditions to generate significant returns while maintaining a favorable risk profile.

Tables 2(a), 2(b), and 2(c) present the results of a back-test simulation from March 2000 through September 2008, which ended in the global financial crises.

Table 2(a) presents the results using the market factor risk model. This model evaluates the performance of the five investment strategies over the GFC period. The market (value-weighted) strategy yielded an average excess return of 7.43%, with a standard deviation of 13.31%, resulting in a Sharpe ratio 0.56. The Equal Weighted strategy demonstrated a higher average excess return of 12.78% and a standard deviation of 14.22%, leading to a Sharpe ratio 0.90. The Minimum Variance strategy showed an average excess return of 10.37%, a lower standard deviation of 8.73%, and a Sharpe ratio 1.19. The Maximum Diversification strategy achieved the highest average excess return of 26.35%, albeit with a higher standard deviation of 16.43% and a Sharpe ratio 1.60. The Risk Parity strategy recorded an average excess return of 12.24%, a standard deviation of 10.91%, and a Sharpe ratio of 1.12, slightly better than the equally weighted portfolio in terms of almost the same return but lower risk. Regarding the effective number of stocks in the portfolios, the risk parity method looks like the equally weighted portfolio with better risk control measures (because of optimization measures). On the other hand, the maximum diversification and minimum variance strategies seek to improve the characteristics of the market portfolio, the former by seeking positions with better alphas and later by holding positions with lower volatilities.

**Table 2(a).** Market Factor Risk Model – GFC (03/2000-09/2008)

|  | Market (Value-Weighted) | Equal Weighted | Minimum Variance | Maximum Diversification | Risk Parity |
|---|---|---|---|---|---|
| Average Excess Return | 7.43% | 12.78% | 10.37% | 26.35% | 12.24% |
| Standard Deviation | 13.31% | 14.22% | 8.73% | 16.43% | 10.91% |
| Sharpe ratio | 0.56 | 0.90 | 1.19 | 1.60 | 1.12 |
| Market Beta | 1.00 | 1.01 | 0.05 | 0.05 | 0.71 |
| Average Positions | 1000.0 | 1000.0 | 87.2 | 94.8 | 1000.0 |
| Effective N | 161.0 | 1000.0 | 52.7 | 54.4 | 761.3 |

The constant correlation covariance matrix showed the highest return during the era that led to the GFC, table 2(b). Because of the systematic consequence of mortgage-backed securities collapse in all market sectors, most major industries (in public markets) were affected in the same direction with delays and different magnitudes. Considering this fact, the maximum diversification portfolio could catch idiosyncratic alphas by assuming the same behavior for all market stocks. An exceptional return of almost 48% during the GFC is proving this. However, because of the market's high volatility in that era, portfolio risk is much higher than other approaches, making it less attractive for asset managers who care more about the Sharpe ratio.

**Table 2(b).** Constant Correlation Covariance Matrix – GFC (03/2000-09/2008)

|  | Market (Value-Weighted) | Equal Weighted | Minimum Variance | Maximum Diversification | Risk Parity |
|---|---|---|---|---|---|
| Average Excess Return | 7.43% | 12.78% | 4.37% | 47.63% | 12.29% |
| Standard Deviation | 13.31% | 14.22% | 8.29% | 59.99% | 13.75% |
| Sharpe ratio | 0.56 | 0.9 | 0.53 | 0.79 | 0.89 |
| Market Beta | 1.00 | 1.01 | 0.32 | 2.65 | 0.98 |
| Average Positions | 1000.0 | 1000.0 | 21.5 | 20.0 | 1000.0 |
| Effective N | 161.0 | 1000.0 | 20.5 | 20.0 | 994.1 |

In Table 2(c), using a shrunk risk model based on the sample covariance matrix, the idiosyncratic risks of the sample are weighted more. As a result, even the maximum diversification strategy has resulted in much less return than it did through the previous risk model. Also, focusing more on individual risk profiles increases effective N by more than 68% in this strategy of the prior risk model. In this risk model, all strategies showed almost similar risk-return profiles. In Figures 2(a), 2(b), and 2(c), the difference between these approaches is clearly shown.

**Table 2(c).** Sample Covariance Matrix with Shrinkage – GFC (03/2000-09/2008)

|  | Market (Value-Weighted) | Equal Weighted | Minimum Variance | Maximum Diversification | Risk Parity |
|---|---|---|---|---|---|
| Average Excess Return | 7.43% | 12.78% | 4.30% | 20.67% | NaN |
| Standard Deviation | 13.31% | 14.22% | 4.34% | 11.47% | NaN |
| Sharpe ratio | 0.56 | 0.9 | 0.99 | 1.8 | NaN |
| Market Beta | 1.00 | 1.01 | 0.25 | 0.70 | NaN |
| Average Positions | 1000.0 | 1000.0 | 146.4 | 54.2 | NaN |
| Effective N | 161.0 | 1000.0 | 83.0 | 33.7 | NaN |

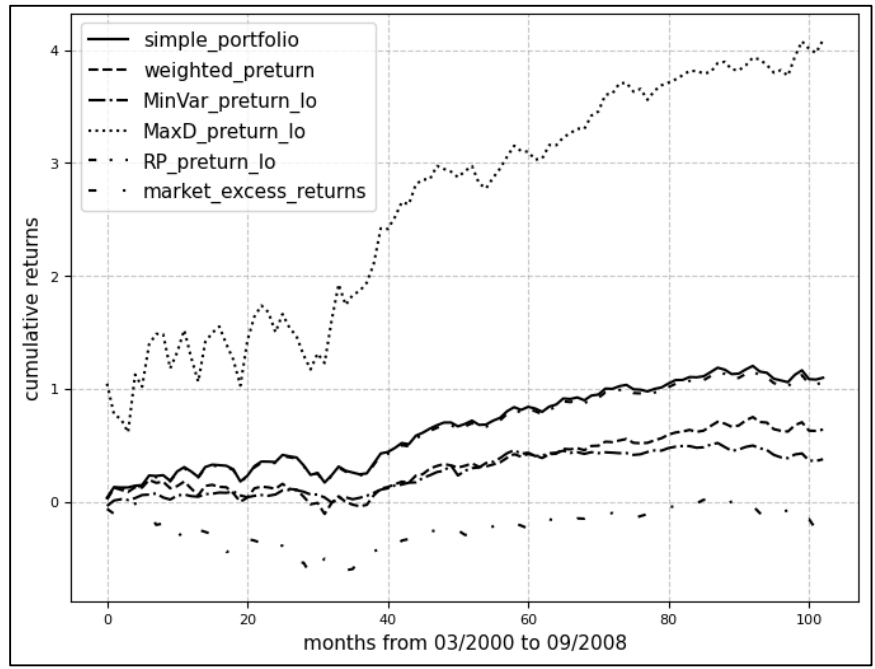

**Figure 2(a).** Constant Correlation Covariance Model – Portfolios Comparison – GFC

Figure 2(a) shows the higher return profile of the maximum diversification portfolio than other investment strategies using a constant correlation covariance matrix. The specific risk profile of this risk model is mostly affected by credit risk and its effects on other industries. Edey (2009) elaborated on the main causes of the crisis, how it has affected the world economy, and how governments and central banks have responded. As stated above, the other risk models showed almost a similar return profile as a result of mainly taking idiosyncratic risk into consideration. Figures 2(b) and 2(c) show the profile and magnitude of market factor and shrunk sample covariance models, respectively.

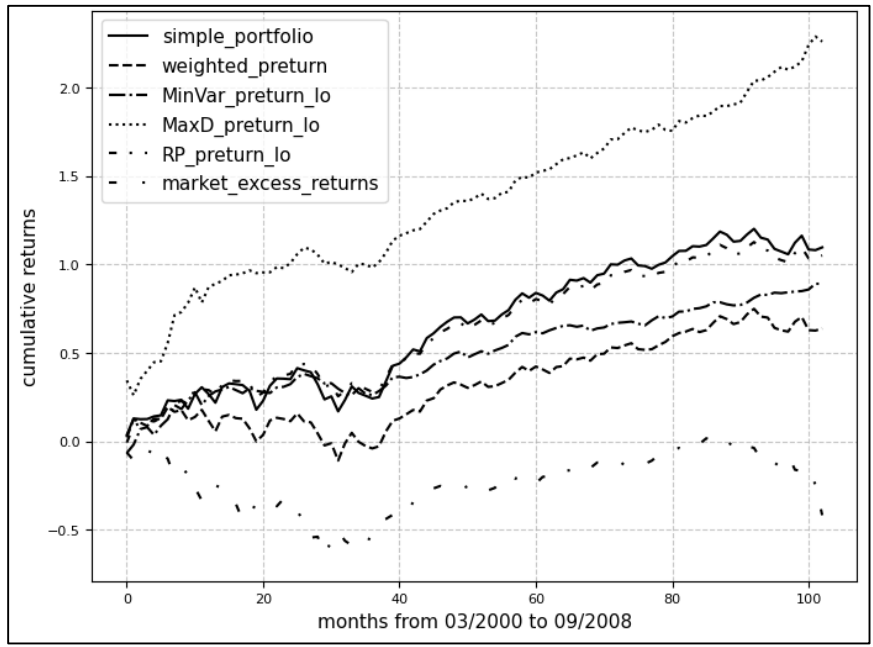

**Figure 2(b).** Market Factor Risk Model – Portfolios Comparison – GFC

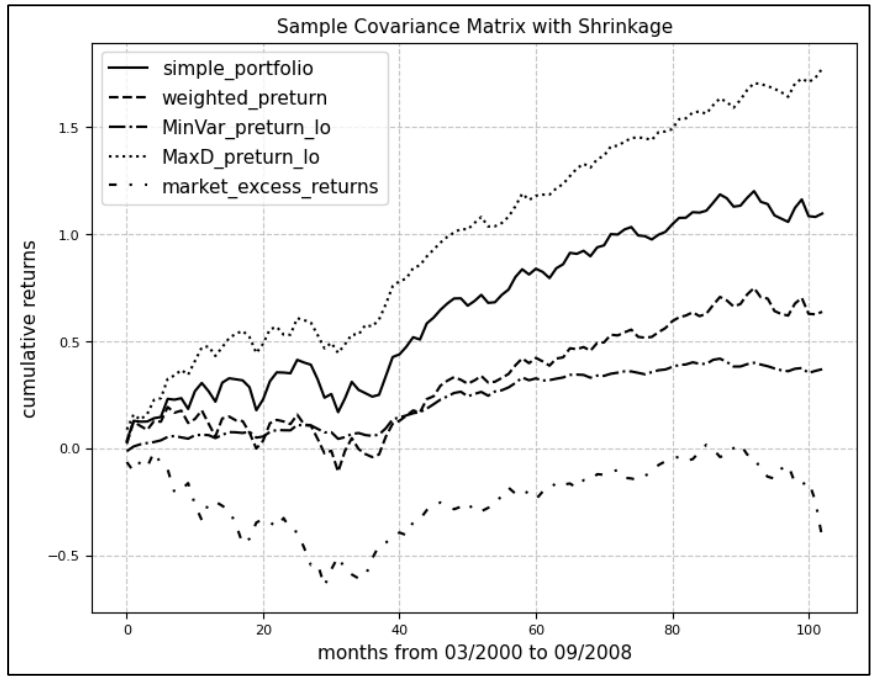

**Figure 2(c).** Shrunk Sample Covariance Matrix – Portfolios Comparison – GFC

Tables 3(a), 3(b), and 3(c) present the results of back-test simulation from September 2008 ending through April 2020 ending in the Covid-19 pandemic using the market factor risk model,

constant correlation covariance matrix, and shrunk sample covariance matrix models, respectively. In terms of the Sharpe ratio, the first and third risk models show similar performance. Minimum variance and maximum diversification portfolios recorded close returns and volatilities. This can be interpreted as similar to the characteristics of both models because of the longer period between the GFC and the COVID-19 crisis. In other words, the sample covariance matrix takes more similar information to the market factor risk model within a more extended period. Figures 3(a) and 3(b) show how similar these models are regarding the shape of plots and volatility. Although the maximum diversification strategy has beaten other portfolios in three risk models, its evolution pattern and final yield using a constant correlation risk model is clearly differentiable from the other portfolio across other risk models, Figure 3(c).

**Table 3(a).** Market Factor Risk Model – Covid-19 (09/2008-04/2020)

|  | Market (Value-Weighted) | Equal Weighted | Minimum Variance | Maximum Diversification | Risk Parity |
|---|---|---|---|---|---|
| Average Excess Return | 11.60% | 11.14% | 4.66% | 22.63% | 10.59% |
| Standard Deviation | 15.28% | 17.31% | 4.80% | 16.64% | 14.01% |
| Sharpe ratio | 0.76 | 0.64 | 0.97 | 1.36 | 0.76 |
| Market Beta | 1.00 | 1.11 | 0.00 | 0.31 | 0.90 |
| Average Positions | 1000.0 | 1000.0 | 39.4 | 51.7 | 1000.0 |
| Effective N | 205.7 | 1000.0 | 25.6 | 33.5 | 831.1 |

**Table 3(b).** Constant Correlation Covariance Matrix – Covid-19 (09/2008-04/2020)

|  | Market (Value-Weighted) | Equal Weighted | Minimum Variance | Maximum Diversification | Risk Parity |
|---|---|---|---|---|---|
| Average Excess Return | 11.60% | 11.14% | 1.61% | 51.26% | 10.52% |
| Standard Deviation | 15.28% | 17.31% | 5.46% | 42.12% | 16.65% |
| Sharpe ratio | 0.76 | 0.64 | 0.29 | 1.22 | 0.63 |
| Market Beta | 1.00 | 1.11 | 0.18 | 2.13 | 1.07 |
| Average Positions | 1000.0 | 1000.0 | 21.0 | 20.1 | 1000.0 |
| Effective N | 205.7 | 1000.0 | 20.4 | 20.0 | 989.8 |

**Table 3(c).** Sample Covariance Matrix with Shrinkage – Covid-19 (09/2008-04/2020)

|  | Market (Value-Weighted) | Equal Weighted | Minimum Variance | Maximum Diversification | Risk Parity |
|---|---|---|---|---|---|
| Average Excess Return | 11.60% | 11.14% | 3.71% | 25.73% | NaN |
| Standard Deviation | 15.28% | 17.31% | 3.97% | 18.68% | NaN |
| Sharpe ratio | 0.76 | 0.64 | 0.93 | 1.38 | NaN |
| Market Beta | 1.0 | 1.11 | 0.21 | 0.98 | NaN |
| Average Positions | 1000.0 | 1000.0 | 111.9 | 39.97 | NaN |
| Effective N | 205.7 | 1000.0 | 54.6 | 27.86 | NaN |

Finally, Tables 4(a), 4(b), and 4(c) present the results of back-test simulation from April 2020 through December 2023, known as the post-Covid-19 era. While this era has not yet led to a financial crisis, its back-test results are presented in this paper just as a comparison between investment portfolios during financial crises and market conditions without major falls.

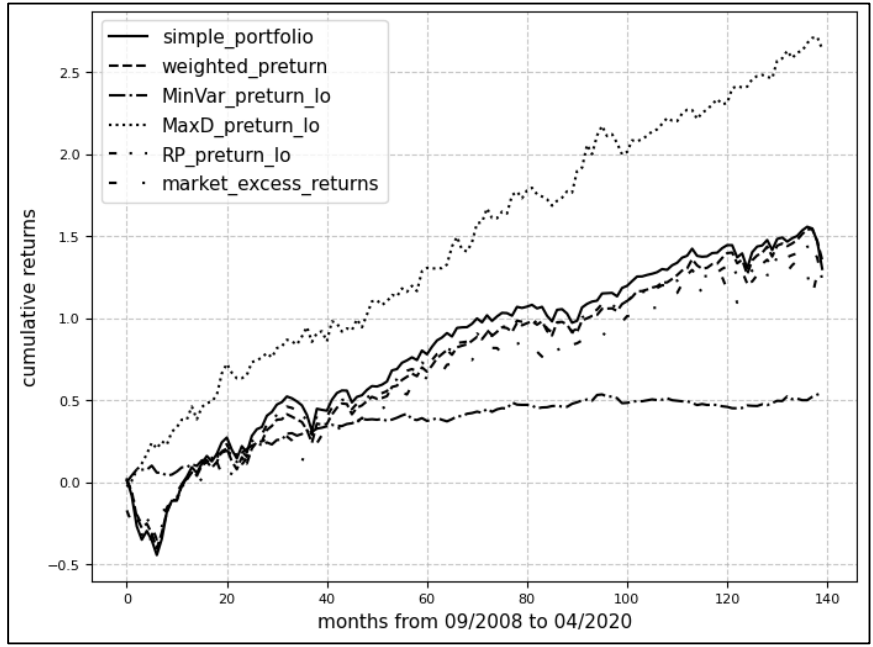

**Figure 3(a).** Market Factor Risk Model – Portfolios Comparison – Covid-19

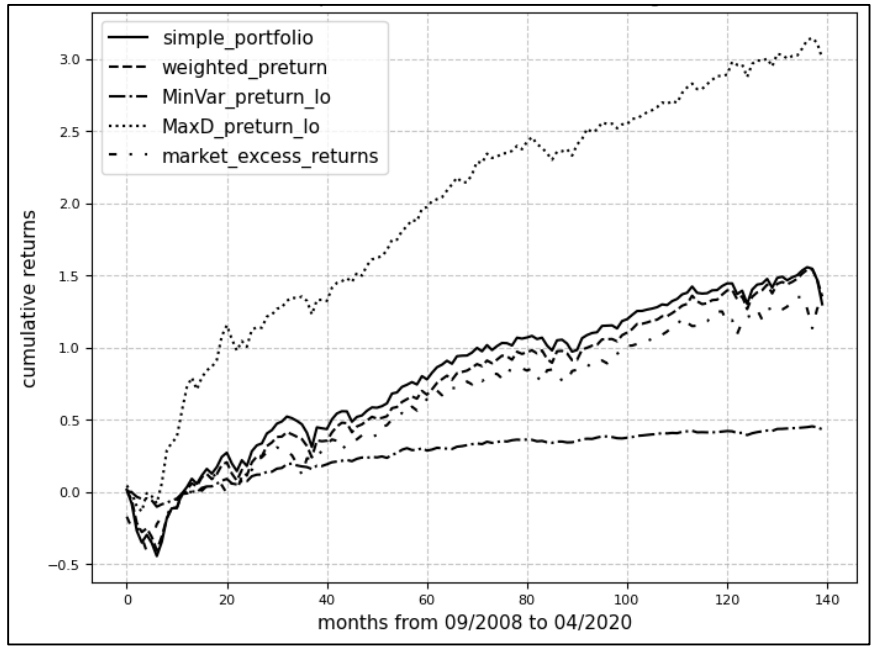

**Figure 3(b).** Shrunk Sample Covariance Matrix – Portfolios Comparison – Covid-

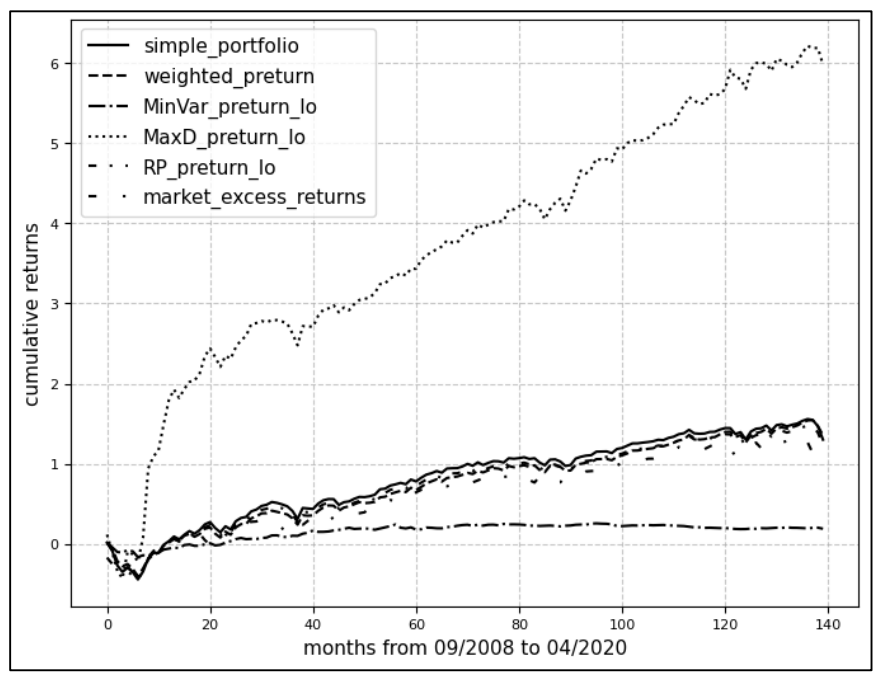

**Figure 3(c).** Constant Correlation Covariance Model – Portfolios Comparison – Covid-19

All the portfolios except minimum variance show increasing excess returns during post-Covid years. For simplicity, we don't include risk parity portfolio analysis for this period. In Table Group 4, the returns of portfolios are shown, which are much higher than in previous eras when dealing with different crises. All risk models have demonstrated similar evolution patterns across portfolios in this period. Still, the maximum diversification strategy can catch more idiosyncratic alphas, which puts it in a much higher rank in excess return. This reward comes with extra risk, which aligns with the mean-variance theory. Also, we realized that the return evolution of the maximum diversification portfolio after COVID-19 follows the reverse pattern recorded between the dot com bubble and GFC periods. While the trend of the market has been going upward, seasonality and significant noises are there due to different reasons, including but not limited to supply chain disruptions after COVID-19, geo-political conflicts influencing the price of assets, especially in oil and gas markets, monetary policies of US federal reserve, etc. However, the current shocks have not been strong enough to create a significant market crash, as had happened with the mentioned crises in the previous decades. Figures 5(a), 5(b), and 5(c) show the 3D evolution of risk, return, and Sharpe ratio over the period from January 1990 through December 2023.

Table 4(a). Market Factor Risk Model – Post Covid (04/2020-12/2023)

|  | Market (Value-Weighted) | Equal Weighted | Minimum Variance | Maximum Diversification | Risk Parity |
|---|---|---|---|---|---|
| Average Excess Return | 18.50% | 15.17% | -3.98% | 49.26% | 12.25% |
| Standard Deviation | 19.69% | 20.88% | 3.23% | 50.36% | 17.53% |
| Sharpe ratio | 0.94 | 0.73 | -1.23 | 0.98 | 0.70 |
| Market Beta | 1.00 | 1.02 | 0.07 | 0.60 | 0.86 |
| Average Positions | 1000.0 | 1000.0 | 26.0 | 38.3 | 1000.0 |
| Effective N | 113.8 | 1000.0 | 21.8 | 27.2 | 842.7 |

Table 4(b). Constant Correlation Covariance Matrix – Post Covid (04/2020-12/2023)

|  | Market (Value-Weighted) | Equal Weighted | Minimum Variance | Maximum Diversification | Risk Parity |
|---|---|---|---|---|---|
| Average Excess Return | 18.50% | 15.17% | -3.38% | 88.39% | 13.66% |
| Standard Deviation | 19.69% | 20.88% | 2.72% | 63.64% | 20.08% |
| Sharpe ratio | 0.94 | 0.73 | -1.49 | 1.39 | 0.68 |
| Market Beta | 1.00 | 1.02 | 0.08 | 1.73 | 0.99 |
| Average Positions | 1000.0 | 1000.0 | 21.3 | 20.3 | 1000.0 |
| Effective N | 113.8 | 1000.0 | 20.5 | 20.1 | 986.4 |

Table 4(c). Sample Covariance Matrix with Shrinkage – Post Covid (04/2020-12/2023)

|  | Market (Value-Weighted) | Equal Weighted | Minimum Variance | Maximum Diversification | Risk Parity |
|---|---|---|---|---|---|
| Average Excess Return | 18.50% | 15.17% | 0.06% | 53.00% | NaN |
| Standard Deviation | 19.69% | 20.88% | 5.67% | 48.37% | NaN |
| Sharpe ratio | 0.94 | 0.73 | 0.01 | 1.10 | NaN |
| Market Beta | 1.00 | 1.02 | 0.23 | 0.92 | NaN |
| Average Positions | 1000.0 | 1000.0 | 229.7 | 37.1 | NaN |
| Effective N | 113.8 | 1000.0 | 121.0 | 25.7 | NaN |

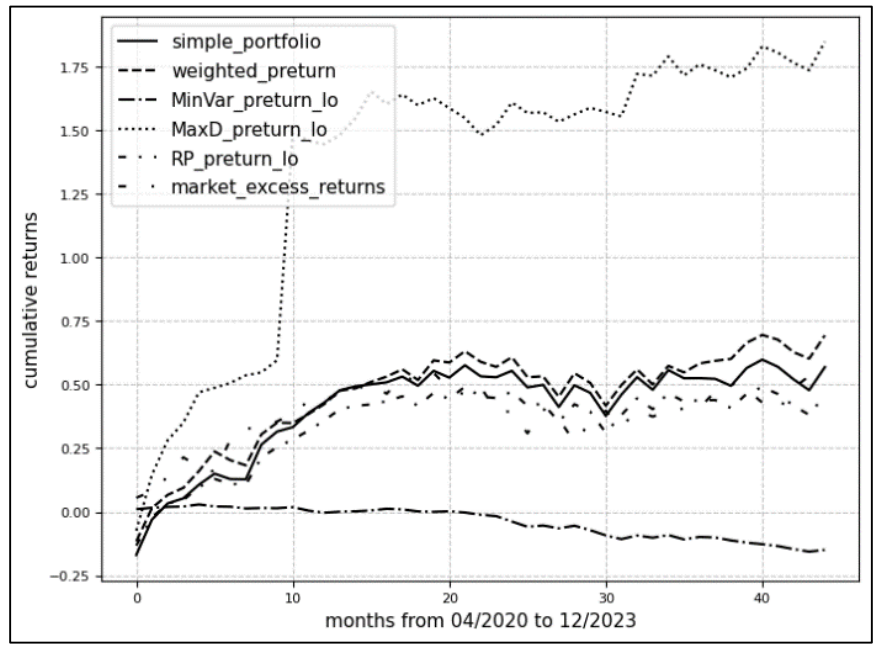

**Figure 4(a).** Market Factor Risk Model – Portfolios Comparison – Post Covid

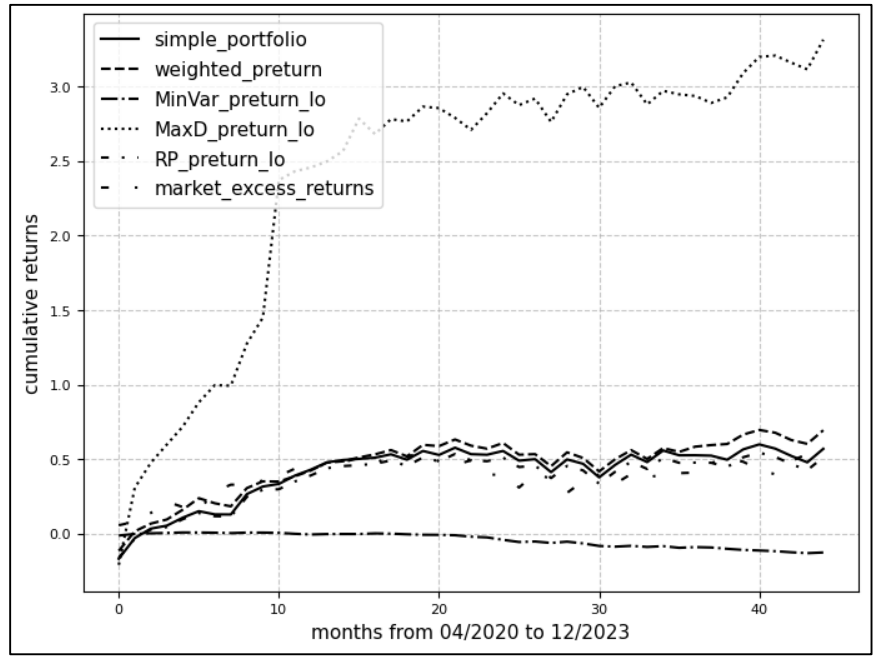

**Figure 4(b).** Constant Correlation Covariance Model – Portfolios Comparison – Post Covid

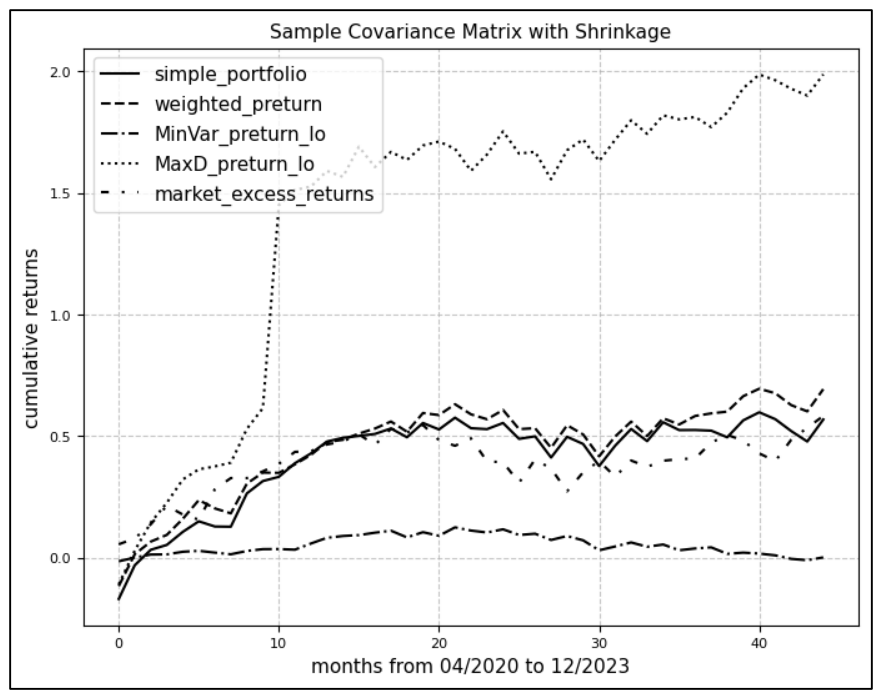

**Figure 4(c).** Shrunk Sample Covariance Model – Portfolios Comparison – Post Covid

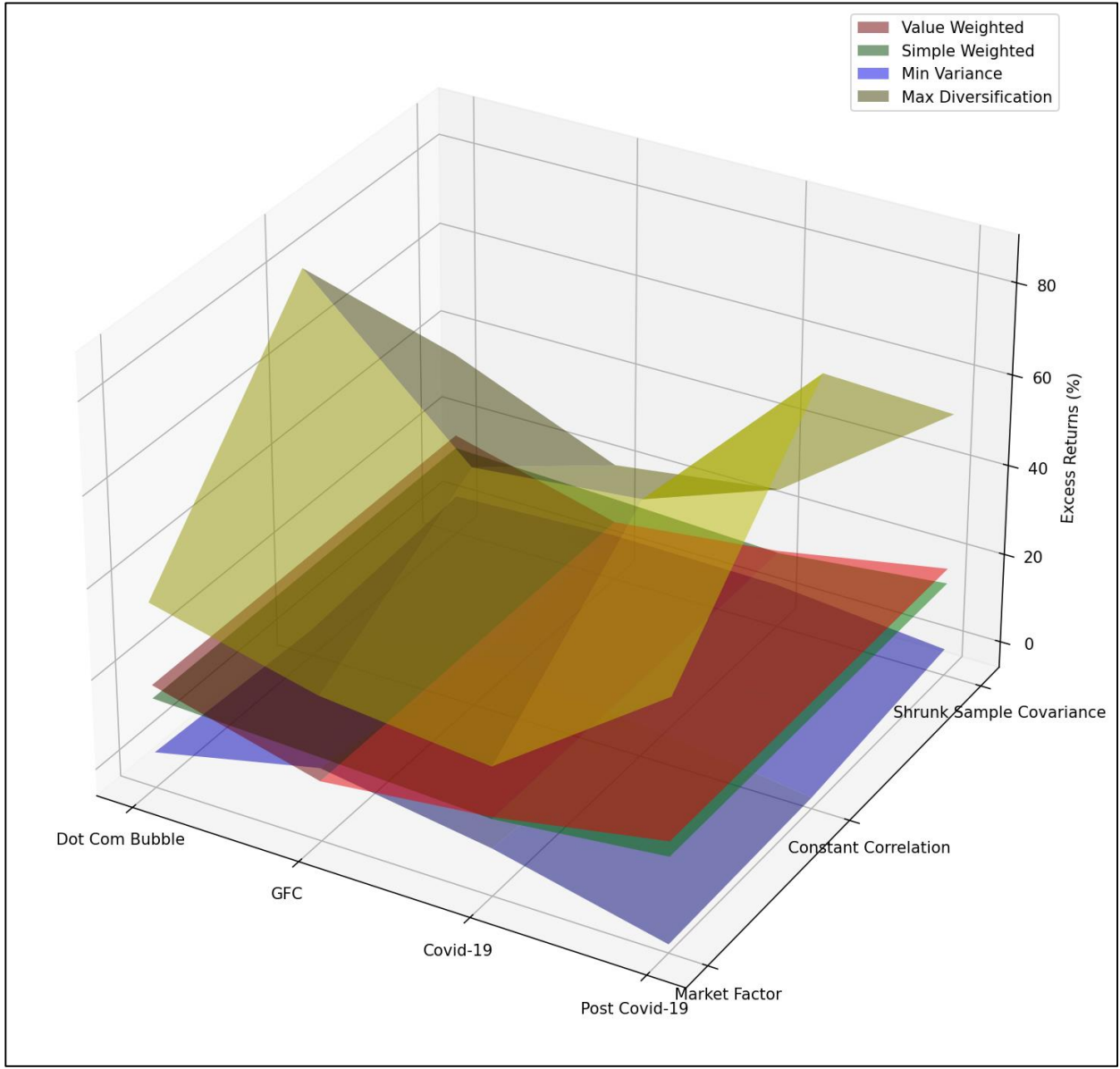

**Figure 5(a).** Excess Return 3D Profiles over Time (%)

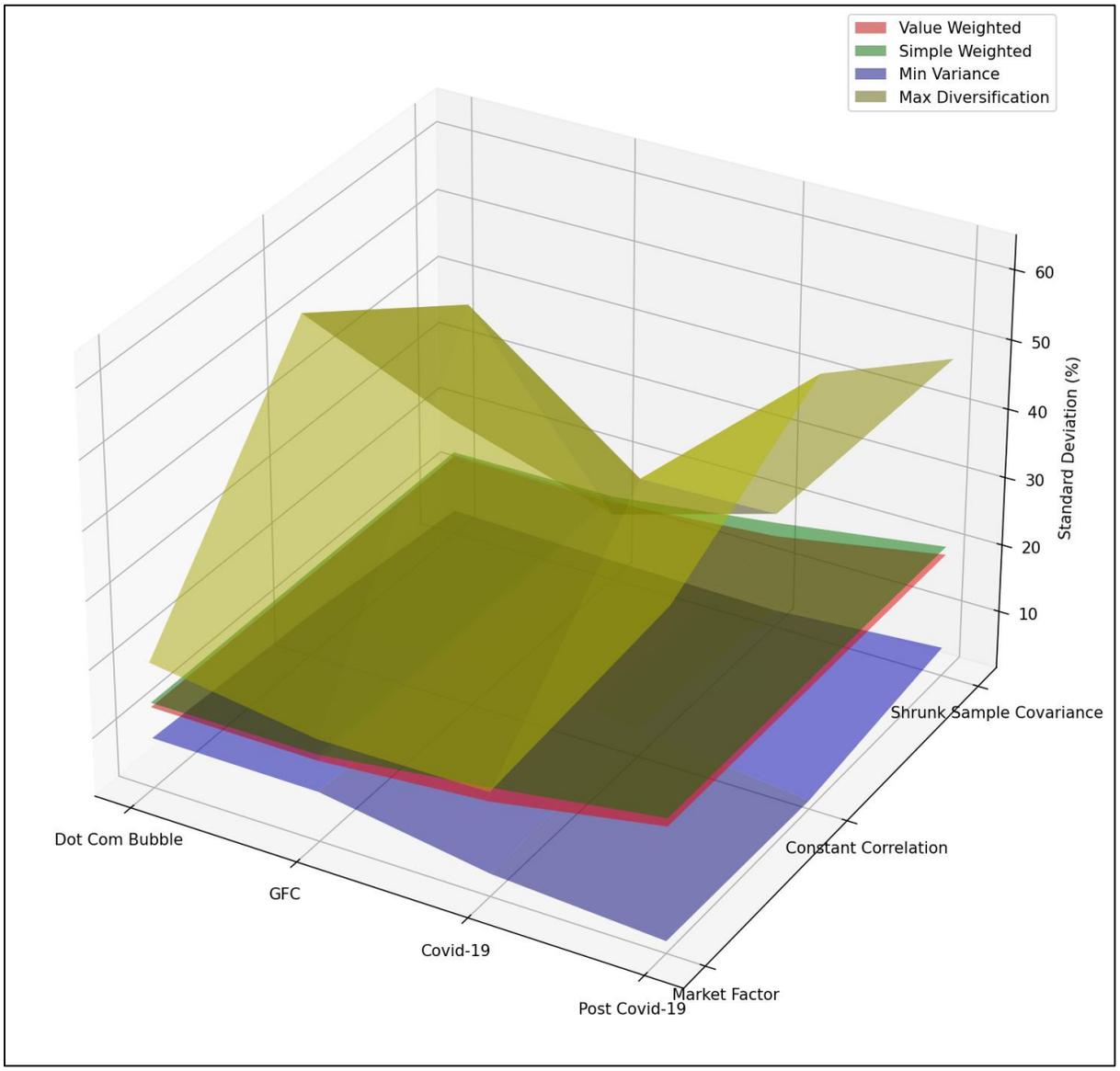

**Figure 5(b).** Standard Deviation 3D Profiles over Time (%)

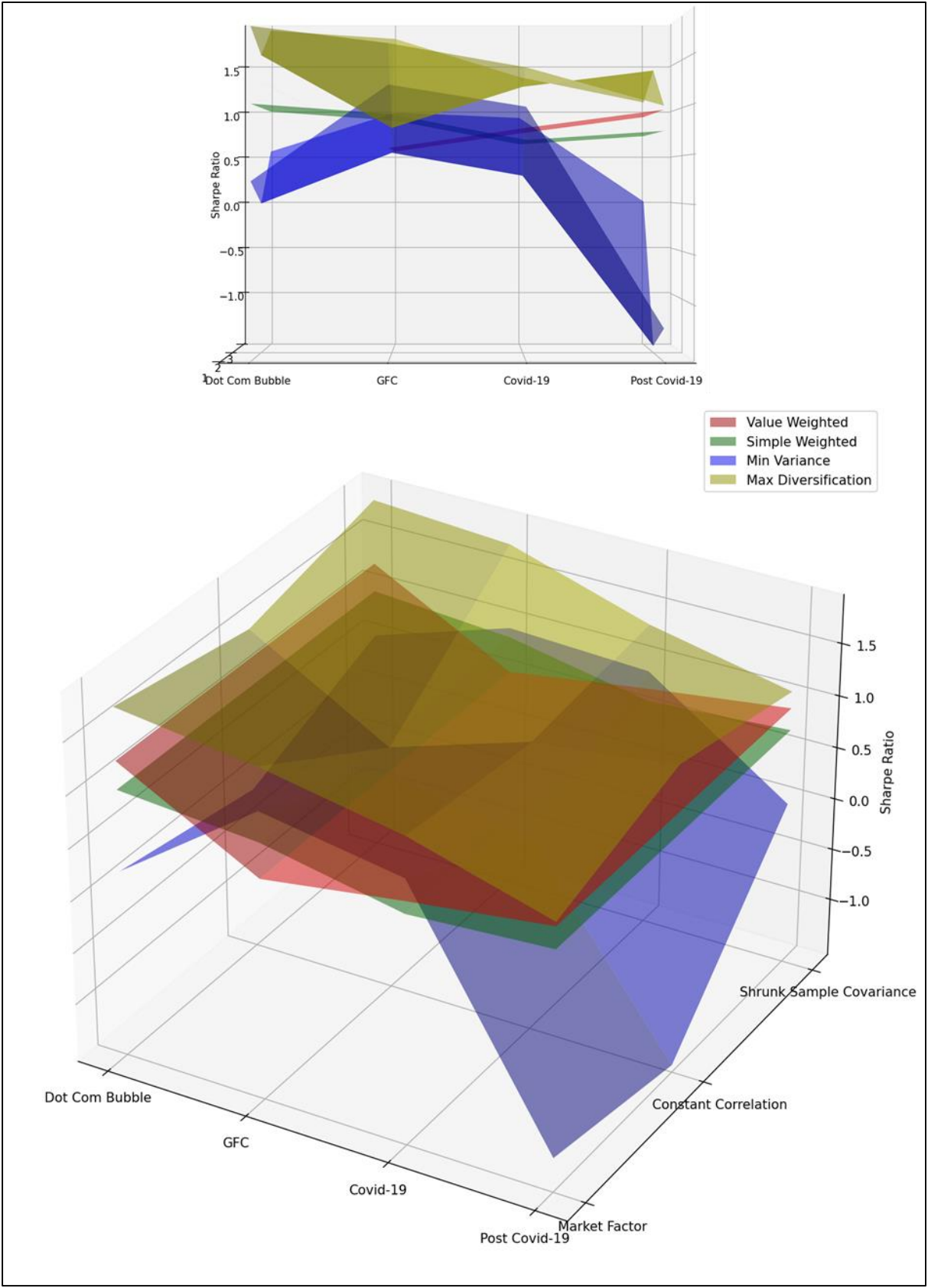

**Figure 5(c).** Sharpe Ratio 3D Profiles over Time

## 4. 1. Managerial Implications

The detailed exploration of investment strategy performances across four distinct market eras—Dot Com Bubble, Global Financial Crisis (GFC), Covid-19 pandemic, and Post-Covid period—using three distinct risk models provides a rich vision for assessing the robustness and adaptability of these strategies under varying economic conditions.

In significant market exuberance during the period of dot-com bubble, the maximum diversification strategy consistently showed high returns across all risk models, highlighting its effectiveness in capitalizing on broad market movements. However, the minimum variance strategy underperformed, indicating its limitations in a rapidly growing market. The market condition during the GFC showed a stark contrast, as more conservative strategies like minimum variance performed relatively better, particularly in terms of the Sharpe ratio under the market factor risk model. This shift underscores the value of defensive strategies in turbulent times. Still, maximum diversification strategies delivered substantial returns under the constant correlation model, albeit at much higher risk levels. The pandemic period again favored the maximum diversification strategy, particularly in the constant correlation model, demonstrating high returns but with increased risk. The minimum variance strategy improved in performance compared to the GFC, reflecting its consistent utility in managing downside risk. Finally, post-pandemic, the maximum diversification strategy under the constant correlation model showed the highest returns, signifying an aggressive but profitable approach during the recovery phase. Conversely, the minimum variance strategy struggled significantly, even yielding negative returns in some instances, possibly due to low market volatility and rising stock correlations in a recovering economy. To understand the strategies' results, we need to gain insight over risk models comparative performance.

Market factor risk model was generally more conservative, with strategies showing lower beta and volatility. It provided a stable framework during the GFC and Post-Covid era, favoring strategies with lower systematic risk. Constant correlation covariance model highlighted the potential for higher returns but also demonstrated considerably higher risks, as evident from the high standard deviations and market betas in the Maximum diversification strategy. It seemed most effective during periods of significant market shifts like the Dot Com and Post-Covid eras. Lastly, sample covariance matrix with shrinkage model offered a more nuanced adaptation to market conditions, often providing a middle ground between the extremes of the other two models. It generally improved the performance of Minimum Variance strategies, enhancing their risk-adjusted returns. The analysis reveals that no single strategy consistently outperforms across all conditions; rather, the choice of strategy should be tailored to the prevailing economic climate and the investor's risk tolerance. Maximum diversification often delivers high returns but comes with significant risk, making it suitable for risk-tolerant investors during market upswings. Conversely, Minimum Variance is preferable for risk-averse investors, particularly during uncertain or declining market conditions.

The comprehensive analysis across different risk models and market eras offers a panoramic view of how different investment strategies can be optimized under various economic conditions. For portfolio managers and investors, the key takeaway is the importance of adaptability and the strategic selection of risk models based on anticipated market dynamics and individual risk profiles. This study not only enhances our understanding of strategic portfolio management but also underscores the critical interplay between market conditions and investment strategy performance, guiding future investment decisions and risk management practices.

## 5. Conclusion

This study underscores the critical importance of adaptability and strategic selection of risk models in portfolio management. By examining investment strategy performances across distinct market eras and using three different risk models, it is evident that no single strategy consistently outperforms across all conditions. The maximum diversification strategy often delivers high returns during market upswings but comes with significant risk, making it suitable for risk-tolerant investors. In contrast, the minimum variance strategy proves more effective for risk-averse investors, particularly during uncertain or declining market conditions. The market factor risk model provided stability during turbulent periods like the GFC, while the constant correlation covariance model highlighted the potential for higher returns in periods of significant market shifts. The sample covariance matrix with the shrinkage model offered a balanced approach, enhancing the risk-adjusted returns of minimum variance strategies. The comprehensive analysis presented in this paper enhances our understanding of strategic portfolio management and emphasizes the interplay between market conditions and investment strategy performance, guiding future investment decisions and risk management practices. Finally, future studies could focus on finding the best numerical modification for risk parity portfolio which is not consistent with the shrunk sample covariance risk model in this research.